\documentclass[11pt,a4paper,fleqn]{article}

\usepackage{graphicx,natbib,amssymb,latexsym,lineno}

\usepackage{amsmath}
\usepackage{amsfonts}
\usepackage{amsthm}

\usepackage{mathrsfs}

\newtheorem{proposition}{Proposition}[section]

\newcommand{\E}{\mathbb{E}}
\newcommand{\Pa}{\mathbb{P}}

\def\T{{ \mathrm{\scriptscriptstyle T} }}

\title{Horvitz--Thompson estimators  for functional data: asymptotic confidence bands and optimal allocation for stratified sampling} 

\author{Herv\'e \textsc{Cardot}, Etienne \textsc{Josserand} \\ 
email : herve.cardot@u-bourgogne.fr, etienne.josserand@u-bourgogne.fr \\
Institut de Math\'ematiques de Bourgogne, UMR CNRS 5584, Universit\'e de Bourgogne \\
 9 Avenue Alain Savary - B.P. 47870, 
21078 DIJON Cedex - France}

\begin{document}
\maketitle

\begin{abstract}
When dealing with very large datasets of functional data, survey sampling approaches are useful in order to obtain estimators of simple functional quantities, without being obliged to store all the data. We propose here a Horvitz--Thompson estimator of the mean trajectory. In the context of a superpopulation framework, we prove under mild regularity conditions  that we obtain uniformly consistent estimators of the mean function and of its variance function. With additional assumptions on the sampling design we state a functional Central Limit Theorem and deduce asymptotic confidence bands. Stratified sampling is studied in detail, and we also obtain a functional version of the usual optimal allocation rule considering a mean variance criterion. These techniques are illustrated by means of a test population of $N=18902$ electricity meters for which we have individual electricity consumption measures every 30 minutes over one week. We show that stratification can substantially improve both the accuracy of the estimators and reduce the width of the global  confidence bands compared to simple random sampling without replacement.
\end{abstract}

\medskip

\noindent \textbf{keywords.}
Asymptotic variance; Functional Central Limit Theorem; 
Superpopulation model; Supremum of Gaussian processes; Survey sampling.

\section{Introduction}

The development of distributed sensors has enabled access to potentially huge databases of signals evolving along time and observed on very fine  scales. Exhaustive collection of such data would require major investments, both for transmission of the signals through networks and for storage. As noted in Chiky \& H\'ebrail (2008), survey sampling of the sensors, which  entails randomly selecting only a part of the curves of the population and which represents a trade off between limited storage capacities and the accuracy of the data, may be relevant compared to signal compression in order to obtain accurate approximations to simple functional quantities such as mean trajectories.

\begin{figure}[h]
\begin{center}
 \includegraphics[height=10cm,width=14cm]{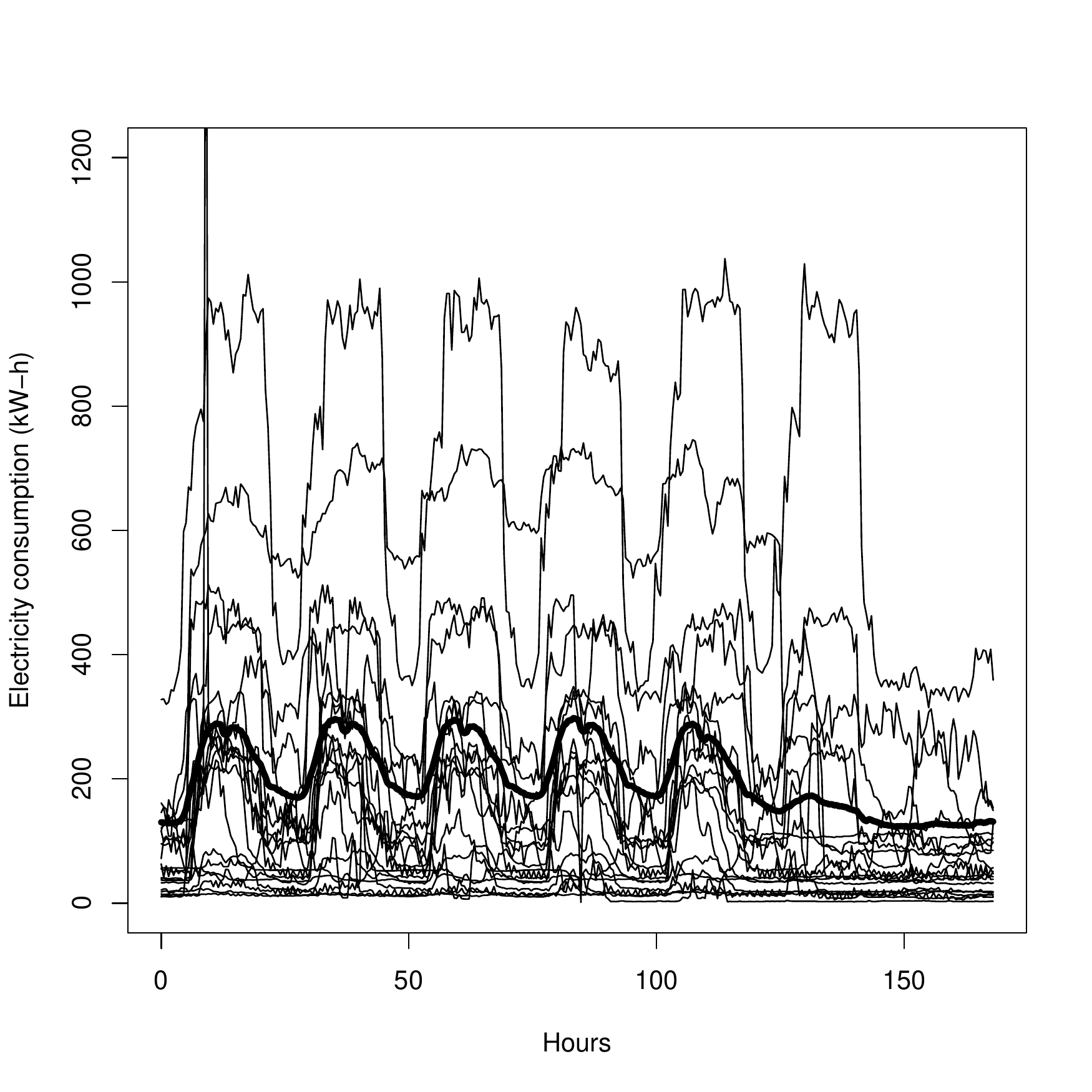}
   \caption{A sample of 20 individual electricity consumption curves. 
   The mean profile is plotted in bold line.}
   \label{fig-meanweek}
   \end{center}
   \end{figure}

Our study is motivated by the estimation, in a fixed time interval, of the mean  electricity consumption curve of a large number of consumers.
The French electricity  operator EDF, \'Electricit\'e De France, intends over the next few years to install over 30 million electricity meters, in each firm and household, which will be able to send individual electricity consumption measures on very fine time scales. Collecting, saving and analyzing all this information, which may be considered as functional, would be very expensive. As an illustrative example, a sample of 20 individual curves, selected among a test population of $N=18902$ electricity meters, is plotted in Figure \ref{fig-meanweek}. The curves consist, for each company selected, of the electricity consumption measured every 30 minutes over a period of one week.  The target is the mean population curve, and we note the high variability between individuals. 

Using survey sampling strategies is one way to get accurate estimates at reasonable cost. The main questions addressed in this paper are to determine the precision of a survey sampling strategy and the strategies likely to improve the sampling selection process in order to obtain estimators that are as accurate as possible and to derive global confidence bands that are as sharp as possible for stratified sampling. There is a vast literature in survey sampling theory ; see for example Fuller (2009). However, as far as we know, the convergence issue with such sampling strategies in finite population has not yet been studied in the functional data analysis literature (Ramsay \& Silverman, 2005,  M\"uller, 2005)  except by Cardot \textit{et al.} (2010), where the objective was to reduce the dimension of the data through functional principal components in the Hilbert space of square integrable functions.
Here we adopt a different point of view and consider the sampled trajectories as elements of the space of continuous functions equipped with the usual sup norm in order to get uniform consistency results through maximal inequalities. Then, it is possible to build  global confidence bands with the help of properties of suprema of Gaussian processes and the functional central limit theorem.

\section{Notation, estimators and basic properties}

Let us consider a finite population $U_N = \{1,\dots,k,\dots,N\}$ of size $N,$ and suppose that to each unit $k$ in $U_N$ we can associate a unique function $Y_k(t),$ for $t \in [0,T],$ with $T<\infty.$
Our target is the mean trajectory
\begin{eqnarray}
 \mu_N(t) &=& \frac{1}{N} \sum_{k \in U} Y_k(t), \quad t \in [0,T].
\label{def:mu}
\end{eqnarray}

We consider a sample $s$ drawn from $U_N$ according to a fixed-size sampling design $p_N(s),$ where $p_N(s)$ is the probability of drawing the sample $s.$ The size $n_N$ of $s$ is nonrandom and we suppose that the first and second order inclusion probabilities satisfy
$\pi_k = \Pa (k \in s) >0$, for all $k \in U_N,$ and $\pi_{kl} = \Pa (k \ \& \ l \in s)>0$ for all $k,l \in U_N$, $k \neq l,$
so that each unit and each pair of units can be drawn with a non null probability from the population. 

It is now possible to write the classical Horvitz--Thompson estimator of the mean curve,
\begin{eqnarray}
   \widehat{\mu}_N(t)  
 & = & \frac{1}{N} \sum_{k \in U} \frac{Y_k(t)}{\pi_k} I_k, \quad t \in [0,T],
\end{eqnarray}
where $I_k$ is the sample membership indicator, $I_k = 1$ if $k \in s$ and $I_k = 0$ otherwise. We clearly have $\E (I_k) = \pi_k$ and $\E (I_k I_l) = \pi_{kl}.$

It is easy to check (Fuller, 2009) that this estimator
 is unbiased, \textit{i.e.} for all $t \in [0,T], $ $\E \{\widehat{\mu}_N (t)\} = \mu_N(t).$ Its covariance function $\gamma_N(s,t) =   \textrm{cov } \{ \widehat{\mu}_N(s), \widehat{\mu}_N(t) \}$ satisfies,  for all $(s,t) \in [0,T]\times [0,T],$
 $$
\gamma_N(s,t) = 
   \frac{1}{N^2} \sum_{k \in U_N} \sum_{l \in U_N}
   \frac{Y_k(s)}{\pi_k} \frac{Y_l(t)}{\pi_l}
   \Delta_{kl},
$$
with $\Delta_{kl} = \pi_{kl} - \pi_k \pi_l$ if $k\neq l$ and $\Delta_{kk} = \pi_k(1-\pi_k).$
An unbiased estimator of $\gamma_N(s,t),$ for all $(s,t) \in [0,T]\times [0,T],$  is
$$
 \widehat{\gamma}_N(s,t) = 
    \frac{1}{N^2} \sum_{k \in s} \sum_{l \in s}
    \frac{Y_k(s)}{\pi_k} \frac{Y_l(t)}{\pi_l}
    \frac{\Delta_{kl}}{\pi_{kl}}.
$$

With real data, such as the electricity consumption trajectories presented in Fig. \ref{fig-meanweek}, we do not observe $Y_k(t)$ at every instant $t$ in $[0,T]$
but only have an evaluation of $Y_k$ at 
$d$ discretization points $0=t_1< \cdots <t_d=T.$ Assuming that there are no measurement errors, which seems realistic in the case of electricity consumption curves,  and that the trajectories are regular enough, linear interpolation is a robust and simple way to obtain accurate approximations of the  trajectories at every instant $t$. For each unit $k$ in the sample $s,$ the interpolated trajectory is defined by
\begin{eqnarray}
  \tilde{Y}_k(t) &=& Y_k(t_i) + \frac{Y_k(t_{i+1})-Y_k(t_i)}{t_{i+1}-t_i}(t-t_i),  \quad t\in[t_i,t_{i+1}].
\end{eqnarray}
It is then possible to define the Horvitz--Thompson estimator of the mean curve based on the discretized observations as
\begin{eqnarray}
\widehat{\mu}_{d}(t) &=& \frac{1}{N} \sum_{k \in s}    \frac{\tilde{Y}_k(t)}{\pi_k},  \quad t \in [0,T].
   \end{eqnarray}
The covariance function of $\widehat{\mu}_d$, denoted by $\gamma_d(s,t) = \textrm{cov } \{\widehat{\mu}_d(s),\widehat{\mu}_d(t)\},$ also satisfies
for all $(s,t) \in [0,T]\times[0,T]$,
\begin{eqnarray}
   \gamma_d(s,t) &=& 
  \frac{1}{N^2} \sum_{k \in U_N} \sum_{l \in U_N}
   \frac{\tilde{Y}_k(s)}{\pi_k} \frac{\tilde{Y}_l(t)}{\pi_l}   \Delta_{kl},
\label{def:thgamma}
\end{eqnarray}
and, as above,  an unbiased estimator of $\gamma_d(s,t)$ is
\begin{eqnarray}
   \widehat{\gamma}_d(s,t) &=& 
    \frac{1}{N^2} \sum_{k \in s} \sum_{l \in s}
    \frac{\tilde{Y}_k(s)}{\pi_k} \frac{\tilde{Y}_l(t)}{\pi_l}  \frac{\Delta_{kl}}{\pi_{kl}} \ .
\end{eqnarray}

To go further we must adopt an asymptotic point of view assuming that the size $N$ of the  population grows to infinity.

\section{Asymptotic Properties}
\subsection{Assumptions}
Let us consider the superpopulation asymptotic framework introduced
by Isaki \& Fuller (1982) and discussed in detail in Fuller (2009).
We consider a sequence of growing and nested populations $U_N$ 
 with size
$N$ tending to infinity
and  a sequence of samples $s_N$ of size $n_N$ drawn from $U_N$ according to the fixed-size sampling designs
$p_{N}(s_N).$ Let us denote by $\pi_{kN}$ and $\pi_{klN}$ their
first and second order inclusion probabilities.
The sequence of sub-populations is an increasing nested one
while the sample sequence is not. For simplicity of notation, we drop
the subscript $N$ in the following when there is no ambiguity.
To prove our asymptotic results, we make the following assumptions. \\

\noindent {\it Assumption} 1.  We assume that  $\displaystyle \lim_{N \rightarrow \infty} \frac{n}{N} = \pi \in ]0,1[.$ \\

\noindent {\it Assumption} 2.  We assume that  $\displaystyle \min_k \pi_k \geq \lambda > 0$,
              $\displaystyle \min_{k \neq l} \pi_{kl} \geq \lambda* > 0$, 
			  $\displaystyle \limsup_{N \rightarrow \infty} \  n \max_{k \neq l} |\pi_{kl} - \pi_k\pi_l| < C_1 < \infty.$ \\
			  
\noindent {\it Assumption} 3.  For all $k \in U,$ $Y_k \in C[0,T],$ the space of continuous functions on $[0,T],$ and $\displaystyle \lim_{N \rightarrow \infty} \mu_N = \mu$
              in $C[0,T].$ \\
              
\noindent {\it Assumption} 4.  There are two positive constants  $C_2$ and $C_3$ and $\beta > 1/2$
            such that,   for all $N$,  $N^{-1} \sum_{k \in U}(Y_k(0))^2 < C_2$ and 
              $N^{-1} \sum_{k \in U}(Y_k(t) - Y_k(s))^2 \leq C_3 |t-s|^{2 \beta}$ for all $(s,t) \in [0,T] \times [0,T].$ \\

Assumptions 1 and 2 concern the moment properties of the sampling designs and are fulfilled  for sampling plans
 such as simple random sampling without replacement or stratified sampling (Robinson \& S\"{a}rndal, 1983, Breidt \& Opsomer, 2000).
Assumptions~3 and 4 are of a functional nature and seem to be  rather weak. Assumption~3 imposes only that the limit of the mean function exists and is continuous, and Assumption~4 states that the trajectories have a uniformly bounded second moment and their mean squared increments satisfy a H\"older  condition.

\subsection{Consistency}

We can now state the first consistency results, assuming that the grid of the $d_N$ discretization points becomes finer and finer  in $[0,T]$ as the population size $N$ tends to infinity. 

 \begin{proposition}\label{prop-muconvergence}
   Let Assumptions 1-4 hold. 
 If  the discretization scheme satisfies \\ $\lim_{N \rightarrow \infty} \max_{ \{i=1, \ldots, d_N-1\}} | t_{i+1}-t_i |^{2 \beta} = o(n^{-1}),$
   then for some constant $C,$
   \begin{displaymath}
\surd{n} \ E \left\{  \sup_{t \in [0,T]} | \widehat{\mu}_d(t) - \mu_N(t) |  \right\} < C.
     \end{displaymath}
  \end{proposition}
Proposition \ref{prop-muconvergence} states that if the grid is fine enough then classical parametric rates of convergence can be attained uniformly, the additional hypothesis meaning that for smoother  trajectories, \textit{i.e.} larger $\beta$, fewer discretization points are needed.
We would also like to obtain that $\widehat{\gamma}_d(t,t)$ is a consistent estimator of the variance function $\gamma_N(t,t)$. To do so, we need to introduce additional assumptions concerning the higher-order inclusion probabilities and the fourth order moments of the trajectories. \\

{\it  Assumption} 5. 
   We assume that \\ $   \lim_{N \rightarrow \infty} \max_{(i_1,i_2,i_3,i_4) \in D_{4,N}}
   \left| E\{ (I_{i_1}I_{i_2}-\pi_{i_1 i_2})
                     (I_{i_3}I_{i_4}-\pi_{i_3 i_4}) \} \right| = 0,
   $ \\
   where $D_{t,N}$ denotes the set of all distinct $t$-tuples
               $(i_1,\dots,i_t)$ from $U_N$. \\
  We also suppose that there are two positive constants  $C_4$ and $C_5,$ such that 
$ N^{-1}\sum_{k \in U_N} Y_k(0)^4 < C_4,$ and
$ N^{-1}\sum_{k \in U_N} \left\{ Y_k(t) - Y_k(s) \right\}^4 < C_5 \left| t-s \right|^{4\beta},$ for all $(s,t) \in [0,T]\times[0,T]$ \\


   The first part  of Assumption 5  is more restrictive than Assumption 2 and is assumed, for example, in  Breidt \&
   Opsomer (2000, part of assumption (A7)). It holds, for instance, in simple random sampling without replacement and stratified sampling. 

  \begin{proposition}\label{prop-gammaconvergence}
  Let Assumptions 1-5 hold.  If  the discretization scheme satisfies  \\ $\lim_{N \rightarrow \infty} \max_{ \{i=1, \ldots, d_N-1\}} | t_{i+1}-t_i |= o(1),$
   then 
$$
   n \ E \left\{  \sup_{t \in [0,T]}  | \widehat{\gamma}_d(t,t) - \gamma_N(t,t) |    \right\} \rightarrow  0, \quad  N \rightarrow \infty.
$$
\end{proposition}
The multiplier $n$ that appears in the Proposition \ref{prop-gammaconvergence} is due to the fact
$n \gamma_N(t,t)$ is a bounded function.
Proposition \ref{prop-gammaconvergence} only states that we can obtain a uniformly consistent estimator of the variance function of the estimated mean trajectory. More restrictive conditions concerning the sampling design would be needed to get rates of convergence.

\subsection{Asymptotic normality and confidence bands}
Proceeding further, we would now like to derive the asymptotic distribution of our estimator $\widehat{\mu}_d$ in order to build asymptotic confidence intervals and bands.
Obtaining the asymptotic normality of estimators in survey sampling is a technical and difficult issue even for simple quantities such as  means or totals of real numbers. Although confidence intervals are commonly used in the survey sampling community, the Central Limit Theorem  has only been checked rigourously, as far as we know, for a few sampling designs.  Erd\"{o}s \& R\'enyi (1959) and H\`ajek (1960) proved that the Horvitz--Thompson estimator is asymptotically Gaussian for simple random sampling without replacement. These results were extended more recently to stratified sampling by Bickel \& Freedman (1994) and some particular cases of two-phase sampling designs by Chen \& Rao (2007). Fuller (2009, \S 1.3) proposes a recent review.
Let us assume that the Horvitz--Thompson estimator satisfies a Central Limit Theorem for real valued quantities with  new moment conditions. \\

{\it Assumption 6}. There is some $\delta>0,$ such that $N^{-1} \sum_{k \in U_N} \left| Y_k(t)\right|^{2+\delta} < \infty$ for all $t \in [0,T],$  and  $\left\{ \gamma_N(t,t) \right\}^{-1/2} \left\{\widehat{\mu}_N(t) - \mu_N(t) \right\}  \rightarrow N(0,1)$ in distribution when $N$ tends to infinity. \\

We can now formulate the following proposition, which tells us that if the sampling design is such that the Horvitz--Thompson estimator of the total of real quantities  is asymptotically Gaussian, then our estimator $\widehat{\mu}_d$ is also asymptotically Gaussian in the space of continuous functions equipped with the sup norm. This means that point-wise normality can be transposed, under regularity assumptions on the trajectories and the asymptotic distance between adjacent discretization points, to a functional Central Limit Theorem.
  \begin{proposition}\label{theo-tclestim}
   Let Assumptions 1-4 and 6 hold and suppose  that the discretization points satisfy \\
   $\lim_{N \rightarrow \infty} \max_{ \{i=1, \ldots, d_N-1\}} | t_{i+1}-t_i |^{2 \beta} = o(n^{-1}).$
   We then have that
\begin{eqnarray*}
   \surd{n} \ \left(\widehat{\mu}_d - \mu_N \right)
   \rightarrow X \ \mbox{ in distribution in } C[0,T]
 \end{eqnarray*}
	where $X$ is a Gaussian random function taking values in $C[0,T]$ with mean $0$ and covariance function $\breve{\gamma}(s,t) = \lim_{N \rightarrow \infty} n \gamma_N(s,t).$
  \end{proposition}
The proof, given in the Appendix, is  based on the Cramer--Wold device which gives access to multivariate normality when considering discretized trajectories. Tightness arguments are then invoked in order to obtain the functional version of the Central Limit Theorem.

Using heuristic arguments similar to those of Degras (2009), we can also build asymptotic confidence bands in order to evaluate the global accuracy of our estimator. To do so, we make use of an asymptotic result from Landau \& Shepp~(1970), which states that  the supremum of a centred Gaussian random function $Z$ taking values in $C[0,T],$ with covariance function $\rho(s,t)$ satisfies
 \begin{eqnarray}
   \lim_{\lambda \rightarrow \infty} \lambda^{-2} \log  \Pa  \left\{ \sup_{t \in [0,T]} Z(t) > \lambda \right\}
  &=&  - \left\{ 2 \textrm{sup}_{t \in [0,T]} \rho(t,t) \right\}^{-1}.
\label{def:LandauShepp}
\end{eqnarray}
Assuming  that $\inf_t \breve{\gamma}(t,t)>0,$ it is easy to prove, with Slutsky's Lemma and Propositions  \ref{prop-gammaconvergence} and   \ref{theo-tclestim}, that the sequence of random functions $Z_n(t)=   \left\{\widehat{\gamma}_d(t,t) \right\}^{-1/2} \left\{\widehat{\mu}_d(t) - \mu_N(t) \right\}$   satisfies the Central Limit Theorem in $C[0,T]$ and converges in distribution to $Z(t).$ Then, the  continuous mapping theorem  tells us that, for each $\lambda >0,$ $\Pa\{\sup_t | Z_n(t)| > \lambda\}$ converges to $\Pa\{ \sup_t | Z(t)| > \lambda\}.$ 
Applying (\ref{def:LandauShepp}) to $Z_n$, a direct computation yields that, for a given risk $\alpha >0$,
  \begin{eqnarray}\label{eq-confidence-bands}
   \Pa \left[ |\widehat{\mu}_d(t)-\mu_N(t)| < \left\{ 2 \log (2/\alpha) \widehat{\gamma}_d(t,t) \right\}^{1/2}, t \in [0,T] \right] & \simeq & 1 - \alpha.
  \end{eqnarray}

Equation  (\ref{eq-confidence-bands}) indicates that, compared to point-wise confidence intervals, global ones can be obtained simply by replacing the scaling given by the quantile of a normal centred unit variance Gaussian variable by the factor $\left\{ 2 \log (2/\alpha) \right\}^{1/2}$. For example, if $\alpha$=0$\cdot$05, respectively  $\alpha$=0$\cdot$01, then $\left\{ 2 \log (2/\alpha) \right\}^{1/2}$= 2$\cdot$716, respectively 3$\cdot$255, instead of 1$\cdot$960, respectively  2$\cdot$576, for a point-wise confidence interval with 0$\cdot$95 confidence, respectively 0$\cdot$99.
The result presented in equation (\ref{def:LandauShepp}) is asymptotic and is therefore more reliable when $\alpha$ is close to zero as seen in our simulation study. 

\section{Stratified sampling designs}

We now consider now the particular case of stratified sampling with simple random sampling without replacement  in all strata, assuming the population $U$ is divided into a fixed number $H$ of strata.
This means that there is a partitioning of $U$ into $H$ sub-populations denoted by   $U_h$, $(h = 1,\dots,H)$.
 We can define the mean curve $\mu_h$ within each stratum $h$ as 
     $  \mu_h(t) = N_h^{-1} \sum_{k \in U_h} Y_k(t), \ t \in [0,T],$
where $N_h$ is the number of units in stratum $h.$ The covariance function, $\gamma_h(s,t),$ within stratum $h$ is defined by $\gamma_h(s,t)  =  N_h^{-1} \sum_{k \in U_h} \{Y_k(s)-\mu_h(s)\} \{Y_k(t)-\mu_h(t)\}, \ (s,t) \in [0,T] \times [0,T].$

In stratified sampling with simple random sampling without replacement in all strata, the first and second order inclusion probabilities are
explicitly known,
and the mean curve estimator of $\mu_N(t)$ is 
 $   \widehat{\mu}_{\textrm{strat}}(t)
    =
    N^{-1} \sum_{h=1}^H n_h^{-1} N_h \sum_{k \in s_h} Y_k(t), \ t \in [0,T], $
where $s_h$ is a sample of size $n_h, $ with $n_h <N_h,$ obtained by simple random sampling without replacement in  stratum $U_h.$
The covariance function of
   $\widehat{\mu}_{\textrm{strat}}$, 
   can  be expressed as
     \begin{eqnarray*}
   \gamma_{\textrm{strat}}(s,t) &= & \frac{1}{N^2}\sum_{h=1}^H N_h \frac{N_h-n_h}{n_h} \ \widetilde{\gamma}_h(s,t), \quad  (s,t) \in [0,T]\times [0,T],
   \end{eqnarray*}
   with  $(N_h-1) \widetilde{\gamma}_h(s,t)  =  N_h \gamma_h(s,t).$

For real valued quantities, optimal allocation rules, which determine the sizes $n_h$ of the samples in all the strata, are generally defined in order to obtain an estimator whose variance is as small as possible.
In our functional context, and as in the multivariate case (Cochran, 1977, \S 5A.2), determining an optimal allocation clearly depends on the criterion to be minimized. 
Indeed, one could consider many different optimization criteria which would lead to different optimal allocations rules.
The width of the global confidence bands derived in equation  (\ref{eq-confidence-bands}) depend only on the standard deviation of the estimator at each instant $t$ and minimising the width at the worst instant of time  or minimizing the average width along time are natural criteria. Nevertheless, finding the solution of such optimization problems is not trivial and not investigated further in this paper. 
If we consider the optimal allocation based on minimising  the mean variance instead of the mean standard deviation, we can then find explicit and simple solutions to
\begin{eqnarray}
   \min_{(n_1, \ldots, n_H)} \int_0^T
  \gamma_{\textrm{strat}}(t,t) \ dt  \  \mbox{ subject to } \ \sum_{h=1}^Hn_h=n \mbox{ and } n_h>0, \ h=1, \ldots, H.
\end{eqnarray}
The solution  is 
   \begin{eqnarray}\label{optimal-allocation}
  n_h^* &=& \displaystyle n \ \frac{N_h S_h}
          {\sum_{i=1}^H N_i S_i }, 
  \end{eqnarray}
with $S^2_h = \int_0^T \widetilde{\gamma}_h(t,t)dt, \ h=1, \ldots, H$,
similar to that of the multivariate case when considering a total variance criterion (Cochran, 1977). This means that a stratum with higher  variance than the others should be sampled at a higher sampling rate $n_h/N_h$. The gain when considering optimal allocation compared to proportional allocation, \textit{i.e.}  $n_h = n N_h/N,$ can also be derived easily.

\section{An illustration with  electricity consumption}
Over the next few years  \'Electricit\'e De France plans to install millions of sophisticated electricity meters that will be able to send, on request, electricity consumption measurements every second. Empirical studies have shown that  even the simplest survey sampling strategies, such as simple random sampling without replacement, are very competitive with signal processing approaches such as wavelet expansions, when the aim is to estimate the mean consumption curve.
To test and compare the different possible strategies, a test population of $N=18902$ electricity meters has been installed in small and large companies. These electricity meters  have read electricity consumption every half an hour over a period of two weeks.

We split the temporal observations and considered only the second week for estimation. The reading from first week were used to build the strata.
Thus, our population of curves is a set of $N=18902$ vectors $Y_k= \{Y_k(t_1), \ldots, Y_k(t_d)\}$ with sizes $d=336$.
Identifying each unit $k$ of the population with its trajectory $Y_k,$ we consider now a particular case of stratified sampling which consists in clustering the space $C[0,T]$ of all possible trajectories  into a fixed number of $H$ strata. 

   \begin{figure}[h]
   \begin{center}
  \includegraphics[height=9cm,width=7.2cm]{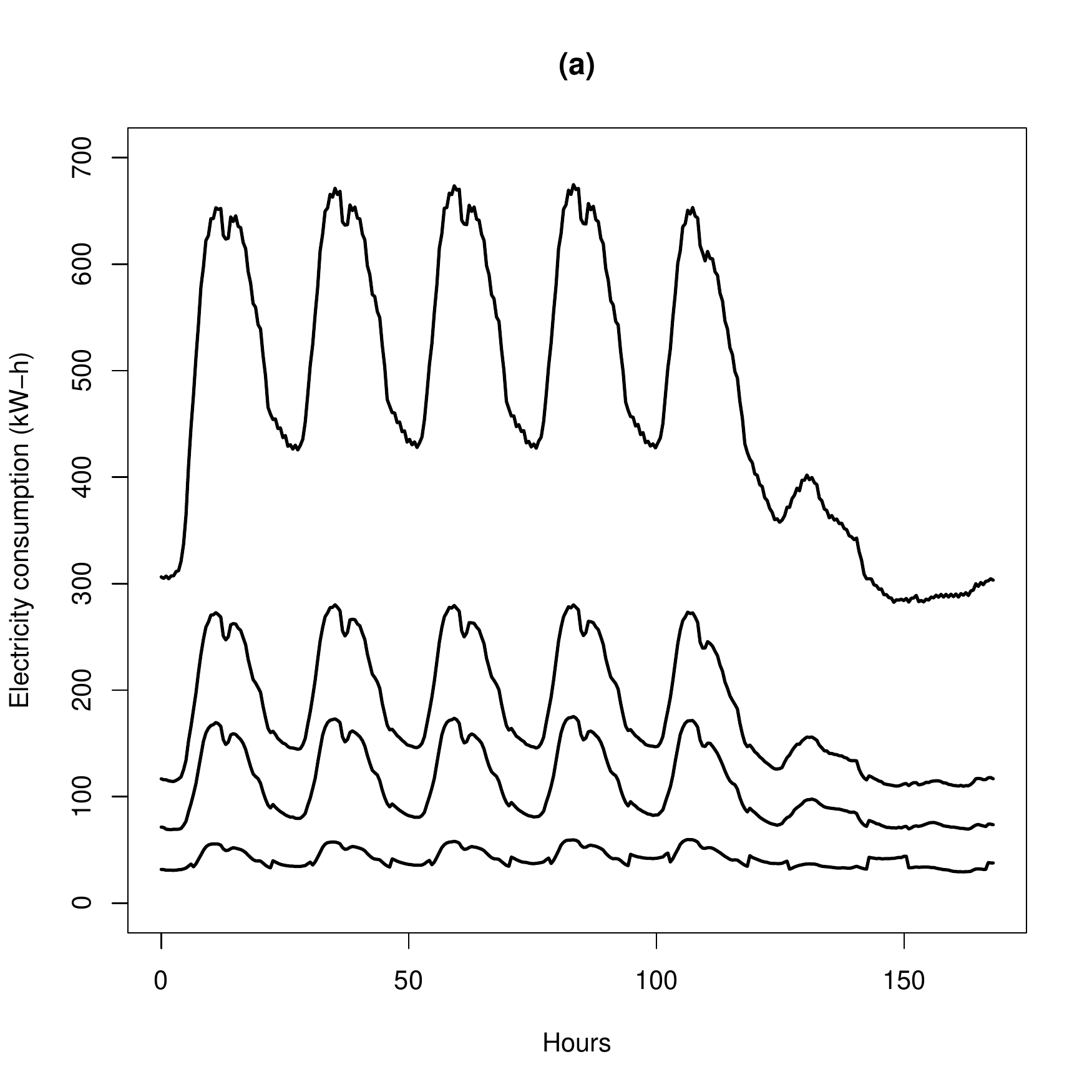}
  \includegraphics[height=9cm,width=7.2cm]{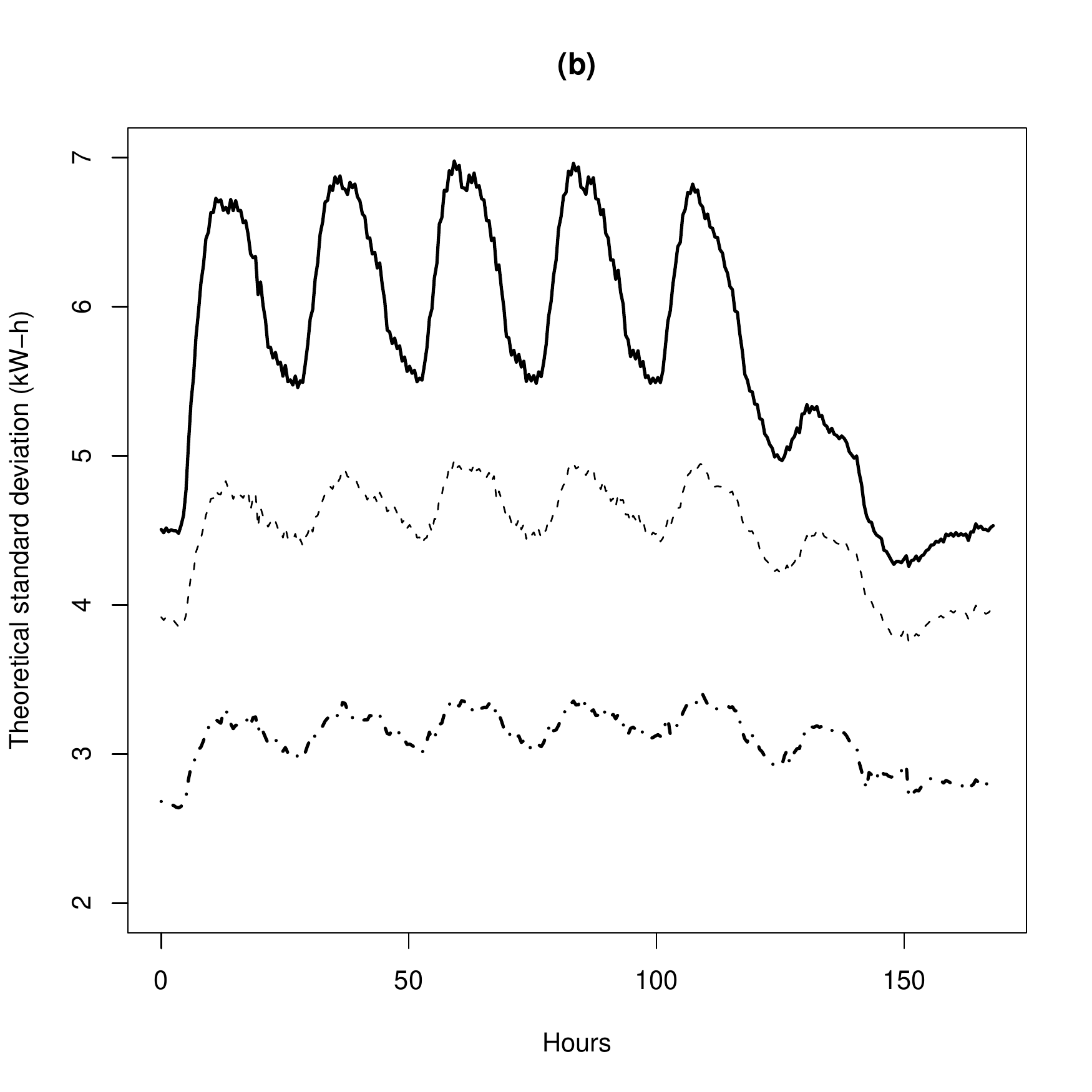}
 \caption{(a) Mean curve in each stratum.  (b) Theoretical standard deviation function $\surd{ \gamma(t,t)}$ for simple random sampling without replacement (solid line), stratified sampling with proportional allocation (dashed line) and stratified sampling with optimal allocation (dotted  dashed line) sampling designs.} \label{fig:strata}
 \end{center}
   \end{figure}

The strata were built by clustering the population according to the maximum level of consumption during the first week.
We decided to retain $H=4$  different clusters based on the quartiles so that all the strata have the same size.
The mean trajectories during the first week in the clusters, drawn in Figure \ref{fig:strata} (a), show a clear size effect. The strata have been numbered  according to global mean consumption. Stratum 4, at the top of Figure \ref{fig:strata} (a), corresponds to consumers  with high global levels of consumption whereas stratum 1, at the bottom of Figure \ref{fig:strata} (a), corresponds to consumers with low global levels of consumption. 

We compared three sampling strategies, with the same sample size $n=2000,$ to estimate the mean population curve $\mu(t)$ and build confidence bands during the second week. In order to evaluate these estimators, we drew 1000 samples using the following sampling designs,
\begin{itemize}
 \item[SRSWR] simple random sampling estimator without replacement, which was first tested by  Electricit\'e de France;
 \item[Proportional] stratified sampling with proportional allocation, in which allocation in each stratum is defined as follows $n_h = n N_h/N$;  the size of each stratum is 500;
\item[Optimal] stratified sampling with optimal allocation according to the rule defined in~(\ref{optimal-allocation}). The sizes of the strata are 126 (stratum 1), 212 (stratum 2), 
 333 (stratum 3) and  1329 (stratum 4).
\end{itemize}

To evaluate the accuracy of the estimators, we considered the following loss criteria, evaluated with discretized data using quadrature rules, for the estimator $\widehat{\mu},$  respectively  $\widehat{\gamma},$ of the mean trajectory, respectively  of the mean variance,
\begin{equation}
   R(\widehat{\mu}) \ = \
         \int_0^T | \widehat{\mu}(t) - \mu(t) |  \ dt,  \quad  
 R(\widehat{\gamma}) \ = \
         \int_0^T | \widehat{\gamma}(t,t) - \gamma(t,t) | \ dt.
           \label{def:muR}
  \end{equation}

Basic statistics for the  estimation errors of the mean function
are given in Table \ref{tab-result-norm}. 
First, we observe that clustering the space of functions by means of stratified sampling leads to a large gain in terms of the accuracy of the estimators. In addition, there is a substantial difference between the proportional and the optimal  allocation rules.
\begin{small}
  \begin{table}[h]
 \caption{Estimation errors for $\mu$ and $\gamma(t,t)$ for the different sampling designs.}{
   \begin{tabular}{rrrrrrrrr}
   \\
   & \multicolumn{4}{c}{Mean function} & \multicolumn{4}{c}{Variance function} \\
    & Mean & 1st quartile & 
    median & 3rd quartile  & mean &  1st quartile & 
    median &  3rd quartile\\
    SRSWR & 4$\cdot$46 & 2$\cdot$37 & 3$\cdot$75 & 5$\cdot$68 & 5$\cdot$26 & 2$\cdot$42 & 4$\cdot$04 & 6$\cdot$64 \\
    Proportional & 3$\cdot$48 & 2$\cdot$03 & 2$\cdot$87 & 4$\cdot$43 & 4$\cdot$77 & 2$\cdot$07 & 3$\cdot$51 & 5$\cdot$79 \\
    Optimal   & 2$\cdot$43 & 1$\cdot$55 & 2$\cdot$10 & 3$\cdot$04 & 1$\cdot$02 & 0$\cdot$56 & 0$\cdot$88 & 1$\cdot$30 \\
   \end{tabular}}
   \label{tab-result-norm}
  \end{table}
\end{small}

We now examine the  true standard deviation functions $\surd{\gamma(t,t)}$, which are proportional to the width of the confidence bands.
They depend on the sampling design and are drawn in Figure \ref{fig:strata} (b). The theoretical standard deviation  is much smaller, at all instants $t,$ for the optimal allocation rule, and it is about  twice smaller   compared to simple random sampling without replacement. There is also a strong periodicity effect in the simple random sampling without replacement due to the lack of control over the units with  high levels of consumption (stratum 4). Estimation errors, according to criterion  (\ref{def:muR}), of the true covariance functions are reported in Table \ref{tab-result-norm}. 
The error is much smaller for stratified optimal allocation than for the other estimators; optimal allocation provides better estimates as well as better estimation of their variance.

Finally, we computed the global confidence bands  to check that formula  (\ref{eq-confidence-bands}), which relies on asymptotic properties of the supremum of Gaussian processes, remains valid when considering confidence levels 0$\cdot$95 and 0$\cdot$99.
The empirical coverage 
is close to the nominal one for  the simple random sampling without replacement, 93$\cdot$8\% and 98$\cdot$3\%, whereas it is a little bit liberal, especially for smaller levels, for the stratified sampling designs, 88$\cdot$7\% and  96$\cdot$8\% for proportional allocation,  and 88$\cdot$1\% and 96$\cdot$8\% for optimal allocation.

\section{Concluding remarks}

The experimental results on a test population of electricity consumption curves confirm that stratification, in conjunction with the optimal allocation rule, can lead, in cases of such high dimensional data, to important gains in terms of the accuracy of the estimation and width of the global confidence bands compared to more basic approaches. We have proposed a simple rule to get confidence bands that could certainly be improved, in terms of empirical coverage, by computing more realistic   scaling factors with  bootstrap procedures (Faraway, 1997) or Gaussian process simulations (Degras, 2010). 

Choosing appropriate strata is also an important aspect of such improvement. Nevertheless, it will generally be impossible to determine for all units  to which cluster they belong. Borrowing ideas from Breidt \& Opsomer~(2008), one possible strategy is to perform clustering on the observed sample and then try to predict to which stratum the units that are not in the sample belong using auxiliary information and supervised classification.

We have assumed that the observed trajectories are not corrupted by  noise at the discretization points. Although this assumption seems quite reasonable in the case of electricity consumption measurements, it is not true in general. Thus, linear interpolation may not always be effective and linear smoother estimators, such a kernels or smoothing splines, would probably be more appropriate ways to obtain functional versions of the discretized observations. 

Finally, another direction for future research is to combine optimal allocation for stratification with model-assisted estimation  when auxiliary information is available. There are close relationships between the shape of electricity consumption curves and variables such as past consumption, temperature, household  area or  type of electricity contract.  Such an estimation procedure relies, as noted in Cardot \textit{et al.}~(2010), on a parsimonious representation of the trajectories in order to reduce the dimension of the data. 
One way to achieve this is to first perform a functional principal components analysis and then to model the relationship between the principal components and the auxiliary information.

\section*{Acknowledgement}
The authors are grateful to the engineers, and particularly to Alain Dessertaine,  of the {\em D\'epartement Recherche et D\'eveloppement} at \'Electricit\'e de France for fruitful discussions and for allowing us to illustrate this research with the electricity consumption data. This article was also improved by comments and suggestions of the referees as well as discussions with Dr. Camelia Goga and Pauline Lardin. Etienne Josserand thanks the {\em Conseil R\'egional de Bourgogne, France} for its financial support (FABER PhD grant). 

\appendix

\section*{Appendix : proofs}

  \begin{proof}[of Proposition~\ref{prop-muconvergence}]
 We study approximation and sampling errors separately :
    \begin{eqnarray}
    \sup_{t \in [0,T]} | \widehat{\mu}_d(t) - \mu_N(t) | & \leq &
       \sup_{t \in [0,T]} | \widehat{\mu}_d(t) - \widehat{\mu}_N(t) | 
     +  \sup_{t \in [0,T]} |   \widehat{\mu}_N(t) - \mu_N(t) | .
     \label{ineq:decompsup}
    \end{eqnarray}
Suppose $t \in [t_i, t_{i+1}[$ then $|Y_k(t)-\tilde{Y}_k(t)| \leq | Y_k(t_i)-Y_k(t_{i+1})| + | Y_k(t)-Y_k(t_i)|.$ By Assumptions 1-2 and an application of the Cauchy--Schwarz inequality,
   \begin{eqnarray}
    | \widehat{\mu}_d(t) - \widehat{\mu}_N(t) | 
    & \leq & \frac{1}{N} \sum_{k \in s} \frac{|Y_k(t)-\tilde{Y}_k(t)|}{\pi_k}   \nonumber \\
   & \leq & \frac{1}{\min_{k \in U_N}\pi_k}  \left[ \frac{1}{N} \sum_{k \in U} \left\{Y_k(t_i)-\tilde Y_k(t) \right\}^2 \right]^{1/2}  \nonumber \\
   & \leq & 
   \frac{1}{\lambda} C_6 |t_{i+1}-t_i |^{\beta}, \nonumber
   \end{eqnarray}
   for some positive constant $C_6$ which does not depend on $t.$ Consequently, 
   \begin{eqnarray}
\surd{n}    \sup_{t \in [0,T]} | \widehat{\mu}_d(t) - \widehat{\mu}_N(t) | & \leq &\surd{n}  \frac{C_6}{\lambda} \max_{i \in \{1, \ldots, d_N-1 \}} |t_{i+1}-t_i |^{\beta} .
    \label{ineq:mud}
   \end{eqnarray}
We now study the sampling error. Consider the pseudo-metric $$d^2_N(s,t) = n E \left\{  \widehat{\mu}_N(t) - \mu_N(t) -\widehat{\mu}_N(s) + \mu_N(s) \right\}^2 $$ for all $(s,t) \in [0,T]\times [0,T].$ We have, for some constant $C_7,$
   \begin{eqnarray}
d^2_N(s,t)  & \leq &
       \frac{n}{N^2} \sum_{k, \ell \in U_N} \left| \frac{\Delta_{k\ell}}{\pi_k\pi_\ell} \right| \left| Y_k(t)-Y_k(s)\right| \left|Y_\ell(t)-Y_\ell(s)\right|  \nonumber \\
    & \leq & \frac{n}{N}  \frac{C_3}{\lambda} |t-s|^{2\beta} 
 + \frac{n}{\lambda^2} \max_{k \neq \ell} | \Delta_{k\ell} | \left[ \frac{1}{N}\sum_{k \in U_N} \left\{Y_k(t) - Y_k(s)\right\}^2\right]
   \nonumber \\
    & \leq & C_7  |t-s|^{2 \beta} . 
    \label{ineq:dist}
   \end{eqnarray}

We apply a result of van der Vaart and Wellner (2000, \S 2.2) based on maximal inequalities to get the uniform convergence and consider the packing number $D(\epsilon, \ d_N)$, which is the maximum number of points in $[0,T]$ whose distance between each pair is strictly larger than $\epsilon.$ It is clear from (\ref{ineq:dist}) that $D(\epsilon, d_N) = O(\epsilon^{-1/\beta}).$ Considering now the particular Orlicz norm with $\psi(x) = x^2$ in Theorem 2.2.4 of van der Vaart and Welner (2000), we directly final that $\int_0^T \psi^{-1}(\epsilon^{-1/\beta}) d\epsilon < \infty$ when $\beta>1/2,$ and consequently there is  a constant $C_8$ such that 
\begin{eqnarray}
E \left\{ \surd{n} \sup_{s,t} |\widehat{\mu}_N(t) - \mu_N(t) -\widehat{\mu}_N(s) + \mu_N(s)| \right\} & \leq & C_8.
\label{ineq:max}
\end{eqnarray}

Since $\sup_t | \widehat{\mu}_N(t) - \mu_N(t)| \leq | \widehat{\mu}_N(0) - \mu_N(0)| + \sup_{s,t} |\widehat{\mu}_N(t) - \mu_N(t) -\widehat{\mu}_N(s) + \mu_N(s) |,$ we get  the announced result with (\ref{ineq:decompsup}), (\ref{ineq:mud}) and 
(\ref{ineq:max}).
  \end{proof}

\begin{proof}[of Proposition~\ref{prop-gammaconvergence}]
 The proof follows the same lines as the proof of  proposition~\ref{prop-muconvergence}. Let us first write, 
    \begin{eqnarray}
    \sup_{t \in [0,T]} | \widehat{\gamma}_d(t,t) - \gamma_N(t,t) | & \leq &
       \sup_{t \in [0,T]} | \widehat{\gamma}_d(t,t) - \widehat{\gamma}_N(t,t) |
     +   \sup_{t \in [0,T]} | \widehat{\gamma}_N(t,t) - \gamma_N(t,t) |
     \label{ineq:decompsup2}
    \end{eqnarray}
Suppose $t \in [t_i, t_{i+1}[$ and define $\delta_{kl}(t) =  | \tilde{Y}_l(t)-Y_l(t) | \  | Y_k(t)|.$
 With Assumptions 1-3, we have, for some constants $C_9$ and $C_{10},$
    \begin{eqnarray*}
 |   \widehat{\gamma}_d(t,t) - \widehat{\gamma}_N(t,t) |
    & \leq &  \frac{C_9}{N^2} \left[ \sum_{k \in s} \left| \tilde{Y}^2_k(t) - Y^2_k(t) \right| +  \max_{k\neq l} | \Delta_{kl} |
    \sum_{k \in s} \sum_{l \neq k}  \left\{ \delta_{kl}(t) + \delta_{lk}(t)\right\}       \right]
  \\
     & \leq & \frac{C_{10}}{N} | t_{i+1} - t_i |^\beta .
    \end{eqnarray*}
 Thus, using Assumption 1,
    \begin{eqnarray}
    n   \sup_{t \in [0,T]} |  \widehat{\gamma}_d(t,t) - \widehat{\gamma}_N(t,t) |& \leq & C_{10} \max_{i \in \{1, \ldots, d_N-1 \}} |t_{i+1}-t_i |^{\beta} .
    \label{ineq:gdisc}
   \end{eqnarray}

Consider now the sampling error and define,   for $(s,t) \in [0,T]\times[0,T],$ $d^2_N(s,t) = n^2  E \left\{  \widehat{\gamma}_N(t,t) - \gamma_N(t,t) - \widehat{\gamma}_N(s,s) +\gamma_N(s,s) \right\}^2$ and $\phi_{kl}(s,t)= Y_{k}(t)Y_l(t)-Y_k(s)Y_l(s).$ We have
    \begin{eqnarray*}
d^2_N(s,t) &  = & 
      \frac{n^2}{N^4} \sum_{k,l \in U_N} \sum_{k',l' \in U_N}
     \phi_{kl}(s,t)   \phi_{k'l'}(s,t) \frac{\Delta_{kl}}{\pi_k \pi_l} \frac{\Delta_{k'l'}}{\pi_{k'} \pi_{l'}}
      E \left\{ \left( \frac{I_k I_l}{\pi_{kl}} -1 \right)  \left( \frac{I_{k'} I_{l'}}{\pi_{k'l'}} -1 \right)\right\} .
    \end{eqnarray*}
Following the same lines as   the proof of Theorem 3 in Breidt \& Opsomer (2000),  we get after some algebra that, for some constant $C_{11},$
\begin{eqnarray}
d^2_N(s,t) & \leq & C_{11} \left[ n^{-1} +  \max_{(k,l,k',l') 
                                     \in D_{4,N}} \left| E \left\{ \left( I_k I_l-\pi_{kl} \right)  \left( I_{k'} I_{l'}-\pi_{k'l'} \right)\right\} \right| \right] | t-s|^{2\beta}.
\end{eqnarray}

Applying again a maximal inequality as in the Proof of Proposition \ref{prop-muconvergence}, we get the announced result.
  \end{proof}

  \begin{proof}[of Proposition~\ref{theo-tclestim}] 
Noting that, with  (\ref{ineq:mud}),
$\surd{n} \left\{ \widehat{\mu}_d(t) - \mu_N(t) \right\} = \surd{n} \left\{\widehat{\mu}_N(t) - \mu_N(t) \right\} + o(1),$
uniformly in $t,$ we only need to study the asymptotic distribution of the random function   
$X_n(t) \ = \ \surd{n}   \left\{ \widehat{\mu}_N(t) - \mu_N(t) \right\},$ for $t \in [0,T].$

   We first consider  a $m$-tuple $(t_1,\dots,t_m) \in [0,T]^m,$ a vector  $c^\T =(c_1,\dots,c_m) \in R^m$ and prove that $ \sum_{i=1}^m c_i X_n(t_i)$ is asymptotically Gaussian for all $c \in R^m.$ Considering  $Y_{kc} =  \sum_{i=1}^m c_i Y_k(t_i),$ it is clear, with Assumption~6, that $N^{-1} \sum_{k \in U} |Y_{kc}|^{2+\delta} < \infty$ and we have   
   \begin{eqnarray*}
   \sum_{i=1}^m c_i X_n(t_i) & = &
    \surd{n} \left\{ \frac{1}{N} \sum_{k \in s}  \frac{Y_{kc}}{\pi_k} 
      -  \sum_{i=1}^m c_i \mu_N(t_i) \right\}.
   \end{eqnarray*}
Denoting by  
$\widehat{\mu}_{c} = N^{-1} \sum_{k \in s} \pi_k^{-1} Y_{kc}$
the Horvitz--Thompson estimator of  $\mu_{c} = N^{-1} \sum_{k \in U} Y_{kc},$  it is clear that  $\mu_{c} = \sum_{i=1}^m c_i \mu_N(t_i),$ 
        $E (\widehat{\mu}_{c}) =  \mu_{c},$ and   with Assumption~6,
   $\surd{n}\big(\widehat{\mu}_{c} 
      - E(\widehat{\mu}_{c}) \big)$
   converges in distribution to $N(0,c^\T Mc)$ where
   $M$ is a covariance matrix with generic elements   $[M]_{ij} = \breve{\gamma}(t_i,t_j)$. The Cramer--Wold device tells us that the vector $(X_n(t_1), \ldots, X_n(t_m))$ is asymptotically multivariate normal.

   Secondly, we need to check that $X_n$ satisfies  a tightness property  in order to get the asymptotic convergence in distribution in the space of continuous functions $C[0,T].$ We have with (\ref{ineq:dist}), for all $(s,t) \in [0,T]\times [0,T],$ 
  $ E \{ | X_n(t)-X_n(s) |^2  \} \leq  C_7 | t-s|^{2 \beta},$
  and  the sequence $X_n$  is tight, when $\beta > 1/2,$ according to Theorem 12.3 of Billingsley (1968).
  \end{proof}

\bibliographystyle{biometrika}

\end{document}